\documentclass[twocolumn,preprintnumbers,superscriptaddress,nofootinbib,aps,prd,floatfix]{revtex4}

\usepackage{amsmath,amssymb}
\usepackage{dsfont}
\usepackage{graphicx} 
\usepackage{epstopdf} 
\usepackage{slashed}
\usepackage{subfigure}
\usepackage{color}
\usepackage{multirow}
\usepackage{pdflscape}

\usepackage{hyperref}

\usepackage{footmisc}

\usepackage{array}
\newcolumntype{L}[1]{>{\raggedright\let\newline\\\arraybackslash\hspace{0pt}}m{#1}}
\newcolumntype{C}[1]{>{\centering\let\newline\\\arraybackslash\hspace{0pt}}m{#1}}
\newcolumntype{R}[1]{>{\raggedleft\let\newline\\\arraybackslash\hspace{0pt}}m{#1}}

\newcommand{\be}{\begin{eqnarray*}}
\newcommand{\ee}{\end{eqnarray*}}

\newcommand{\bee}{\begin{eqnarray}}
\newcommand{\eee}{\end{eqnarray}}
\newcommand{\beeq}{\begin{equation}}
\newcommand{\eeeq}{\end{equation}}

\newcommand{\totwo}{$\times 10^{-2}$}
\newcommand{\tothree}{$\times 10^{-3}$}
\newcommand{\tofour}{$\times 10^{-4}$}

\begin{document}

\title{Di-Higgs phenomenology: The forgotten channel}
\begin{abstract} 
  Searches for multi-Higgs final states allow to constrain parameters
  of the SM (or extensions thereof) that directly relate to the
  mechanism of electroweak symmetry breaking. Multi-Higgs production
  cross sections, however, are small and the phenomenologically
  accessible final states are challenging to isolate in the busy
  multi-jet hadron collider environment of the LHC run 2. This makes
  the necessity to extend the list of potentially observable
  production mechanisms obvious. Most of the phenomenological analyses
  in the past have focused on $gg\to hh+jets$; in this paper we study
  $pp\to t\bar t hh$ at LHC run 2 and find that this channel for
  $h\to b\bar b$ and semi-leptonic and hadronic top decays has the
  potential to provide an additional handle to constrain the Higgs
  trilinear coupling in a global fit at the end of run 2.
\end{abstract}
%
%
\author{Christoph Englert} \email{christoph.englert@glasgow.ac.uk}
\affiliation{SUPA, School of Physics and Astronomy, University of
  Glasgow,\\Glasgow, G12 8QQ, United Kingdom}

\author{Frank Krauss} \email{frank.krauss@durham.ac.uk}
\affiliation{Institute for Particle Physics Phenomenology, Department
  of Physics,\\Durham University, DH1 3LE, United Kingdom}

\author{Michael Spannowsky} \email{michael.spannowsky@durham.ac.uk}
\affiliation{Institute for Particle Physics Phenomenology, Department
  of Physics,\\Durham University, DH1 3LE, United Kingdom}

\author{Jennifer Thompson} \email{jennifer.thompson@durham.ac.uk}
\affiliation{Institute for Particle Physics Phenomenology, Department
  of Physics,\\Durham University, DH1 3LE, United Kingdom}

\preprint{IPPP/14/82, DCPT/14/164}

\maketitle

\section{Introduction}
The Higgs discovery in 2012 by the ATLAS and CMS
experiments~\cite{hatlas,hcms} and subsequent preliminary comparisons
of its properties against the Standard Model (SM)
expectation~\cite{Hcoup} have highlighted its SM character in standard
measurements. The next step in demystifying the nature of the
electroweak scale will therefore crucially rely on precise
measurements of the Higgs' properties at low as well as high momentum
transfers during run 2, and on constraining or even measuring Higgs
properties that have not been in the sensitivity reach during run 1.

A parameter in the SM that is directly sensitive to spontaneous
symmetry breaking is the quartic Higgs coupling~$\eta$
\begin{multline}
  \label{eq:hpot}
  V(H^\dagger H) = \mu^2 H^\dagger H + {\eta \over 2} (H^\dagger H)^2\\
  \supset {1\over 2} m_h^2h^2 + {\sqrt{ {\eta} }\over 2}m_h h^3 +
  {\eta\over 8}h^4 \,,
\end{multline}
where we use the unitary gauge $H^T=(0,(v+h)/\sqrt{2})$ and $v\simeq
246~\text{GeV}$. The second independent parameter in the SM Higgs
potential $\mu^2<0$ is reverse--\-engineered to obtain an acceptably
large value of the electroweak symmetry breaking scale and pole mass
\begin{equation}
  \label{eq:smcorr}
  (173~\text{GeV})^2\simeq {v^2\over 2}= {-\mu^2 \over \eta}\,, 
  \quad m_h^2=\eta v^2 
\end{equation}
for a given Higgs self-coupling $\eta$. These relations determine a
unique value of the Higgs self-coupling in the SM $\eta={m_h^2/ v^2}$
as required by renormalisablility. 

To obtain a measurement of the Higgs self-coupling $\eta$, we may
think of Eq.~\eqref{eq:hpot} as the lowest order in an effective field
theory expansion in a new physics scale $\Lambda$.  A new operator possibly
relevant for softening the correlation of Higgs mass and electroweak
scale is, e.g., $O_6=(H^\dagger H)^3$.  Consequently, in the absence of 
additional new resonant phenomena related to electroweak symmetry breaking 
and in order to prove or disprove the existence of such operators, a question 
that needs to be addressed is how well the Higgs self-interaction 
parameter can be constrained assuming the standard low-energy Higgs 
phenomenology only. 

Our best option to phenomenologically access the relevant parameter
$\eta$ at the LHC is via its impact on di-Higgs
production~\cite{nigel,contino} via the trilinear Higgs
self-coupling. Inclusive di-Higgs cross sections typically have cross
sections in the ${\cal{O}}$(10~fb)
range~\cite{Frederix:2014hta,Baglio:2012np}.  This implies that, in order 
to analyse them, the large SM-like Higgs branching ratios $h\to b\bar
b,\tau^+\tau^-$~\cite{us,us2,danilo} and 
$h\to W^+W^-$~\cite{Papaefstathiou:2012qe} must be employed.  Advanced 
substructure techniques~\cite{htagging,shower_dec} or small irreducible 
backgrounds such as in $hh\to b\bar b \gamma \gamma$~\cite{uli,atlasuli}
are crucial in most analyses to date, which have focused on the dominant 
di-Higgs production cross section, gluon fusion (GF) with 
$\sigma^{\text{NLO}}\simeq
30~\text{fb}$ \cite{nloqcd}.  To increase sensitivities in this channel 
emission of an additional jet has been discussed in Refs.~\cite{us,andreas}; 
a complete analysis of WBF-like production in $pp\to hhjj$ has become 
available only recently~\cite{Dolan:2013rja}.

Common to all realistic di-Higgs analyses discussed in the literature
is that they will be sensitive to systematic uncertainties at the end
of run 2 and it is quite likely that measurements in only a single
di-Higgs channel will not provide enough information to formulate a
significant constraint on the Higgs self-interaction in the above
sense~\cite{jose2}. Hence, it is mandatory to extend the list of
potential phenomenologically interesting search channels in
proof-of-principle analyses.

In this paper we investigate $pp\to t\bar t hh$, and study
semi-leptonic and hadronic top decays $t\to \ell\nu b$ and $h\to b\bar
b$. In particular, we discuss the phenomenological appeal of this
particular di-Higgs final state as a function of the number of applied
$b$-tags. We first study the qualitative behaviour of $pp\to t\bar
thh$ in Sec.~\ref{sec:signal}, where we also comment on the signal and
background event generation employed in the remainder of this work. In
Sec.~\ref{sec:analysis}, we detail our analysis and we discuss the
sensitivity of $pp\to t\bar t hh$ to in detail before we present our
conclusions in Sec.~\ref{sec:conc}.

\section{Signal Cross Section Sensitivity and Event Generation}
\label{sec:signal}
The sensitivity of di-Higgs cross sections from GF and WBF is dominated
by destructive interference of the continuum production and the
subamplitude containing proportional to the trilinear coupling
$\lambda$. In gluon fusion this is apparent from low-energy effective
theory arguments~\cite{Kniehl:1995tn} by expanding 
\begin{multline}
  {\cal L}_{\text{LET}} = -{\alpha_s \over 12 \pi} G^a_{\mu\nu} G^{a\,
    \mu\nu} \log \left( 1+ {h\over v}\right) \\= {\alpha_s \over 12
    \pi} G^a_{\mu\nu} G^{a\, \mu\nu} \left( {h^2\over 2 v^2} - {h\over
      v} \right)\,,
\end{multline}
which makes the relative minus between the continuum and the $gg\to
h\to hh$ diagrams explicit. As a consequence, the gluon fusion cross
section is a decreasing function with
$\lambda\gtrsim\lambda_{\text{SM}}$. In WBF the destructive character
is explicit from nested cancellations that are similar to
unitarity-based cancellations observed in longitudinal gauge boson
scattering.

\begin{figure}[!t]
  \includegraphics[width=0.42\textwidth]{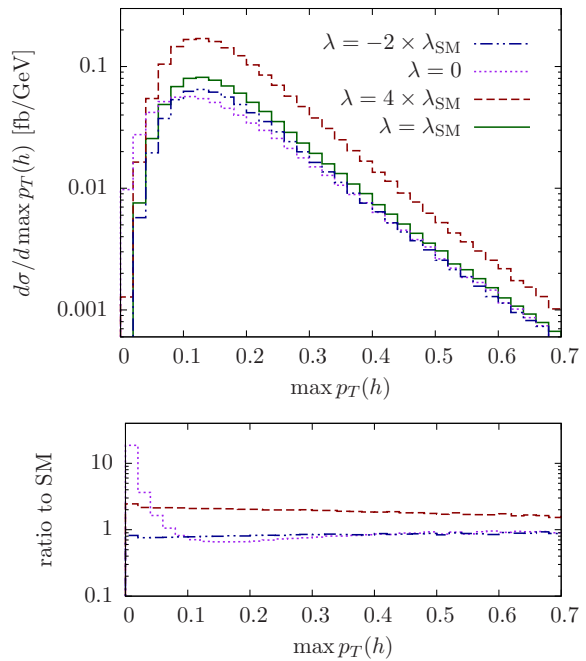}
  \caption{\label{fig:pt} Differential distributions at 14~TeV centre
    of mass energy of the inclusive maximum Higgs transverse momentum
    for different values of the Higgs trilinear coupling
    $\lambda$. The lower panel displays the ratio of the $\max p_T(h)$
    distribution with respect to the SM
    ($\lambda=\lambda_{\text{SM}}$).}
\end{figure}

Qualitatively different from GF- and WBF-induced di-Higgs production,
$pp\to t\bar t hh$ is the leading cross section which is impacted by
{\it constructive} interference, yielding an increasing cross section
with $\lambda > \lambda_{\text{SM}}$, Fig.~\ref{fig:pt}. Quite
different to loop-induced gluon fusion di-Higgs production, there is
no characteristic threshold scale involved in $pp\to t \bar t hh$ that
can be exploited in a targeted boosted search strategy~\cite{us,us2};
the $t\bar t hh$ cross section is a rather flat function of
$\lambda$~\cite{Frederix:2014hta} and differential distributions away
from production threshold do not show a significant deviation apart
from a global rescaling of the differential distribution by
$\sigma(\lambda\neq \lambda_{\text{SM}})/\sigma(\lambda_{\text{SM}})$
for a transverse momentum range that is interesting for the
experiments (Fig.~\ref{fig:pt}). Furthermore, the expected inclusive
$t\bar t hh$ cross section with $\sigma\simeq 1~\text{fb}$ at a 14 TeV
LHC asks for a selection as inclusive as possible to be sensitive to
the signal contribution even for a target luminosity of 3/ab in the
first place.

If we treat the top-Yukawa interaction as legacy measurement and set
$y_t=y_t^{\text{SM}}$, we can imagine a physics situation with an
enhanced trilinear coupling that renders the dominant gluon fusion
modes suppressed but leaves an excess in $pp\to t\bar t hh$
production. In the general dimension six extension alluded to in the
introduction this corresponds to a negative Wilson coefficient of
$O_6$. Enhanced Higgs self-couplings have been discussed more
concretely in the context of conformal Coleman-Weinberg-type
extensions of the SM in~\cite{Abel:2013mya}. Obviously, the opposite
phenomenological situation of $\lambda < \lambda_{\text{SM}}$ is
accompanied by enhanced GF and WBF di-Higgs cross sections while
$pp\to t\bar{t} hh$ becomes smaller (however the cross section becomes
rather flat). Such a situation occurs for instance in composite Higgs
scenarios~\cite{contino}, which typically have a smaller Higgs
trilinear coupling than predicted in the SM (in addition to modified
top Yukawa interactions). Therefore, comparing the measured rates and
(ideally) distributions in all three channels, i.e. gluon-fusion,
weak-boson-fusion and in association with a top quark pair, provides a
precision tool for BSM electroweak symmetry breaking.

  \begin{table*}[!t]
    \begin{center}
      \begin{tabular}{|| r || c | c  || c |c |c |c |c |c || }
        \hline
        & \multicolumn{2}{|c||}{signal} &
        \multicolumn{6}{c||}{backgrounds} \\
        \hline
        &$\xi=1$ & $\xi=4$ & 
        $t\bar t b\bar b b\bar b$  & $t\bar t h b\bar b$ & $t\bar t hZ$ & $t\bar t Z b\bar b$ &
        $t\bar t ZZ$ & $Wb\bar b b\bar b$ \\
        \hline
trigger                            &  0.10                 &  0.23                  &  4.75                 &  1.38                &  0.64                & 1.37                   &  1.36\totwo                  &   1.33  \\
jet cuts                           & 7.40\totwo        &  0.17                  &  1.44                &  0.76                &  0.40                 & 0.65                  & 8.74\tothree                  &  7.46\totwo \\
5 $b$ tags                     &  1.23\totwo      &  2.83\totwo        &  4.46\totwo     &  6.19\totwo     &  7.24\tothree    & 4.43\totwo       & 1.25\tothree         & 5.35\tofour\\
$2 \times h\to b\bar b$               &  7.33\tothree   & 1.69\totwo        &  1.59\totwo      & 2.71\totwo      &  3.41\tothree   & 1.56\totwo       & 4.28\tofour         &  $<$1\tofour\\
lep./had. $t$                  &  5.04\tothree   & 1.12\totwo       &  9.50\tothree   & 1.66\totwo      &  2.29\tothree   & 9.42\tothree     & 2.69\tofour        &   $<$1\tofour\\
lep. $t$ only                          & 2.33\tothree    & 5.29\tothree    &  5.03\tothree    & 9.36\tothree   &  1.14\tothree   & 4.90\tothree     & 1.39\tofour     &   $<$1\tofour\\ 
had. $t$ only                         &  2.71\tothree   & 5.93\tothree       &  4.47\tothree    &  7.20\tothree  &  1.16\tothree   &  4.44\tothree     &  1.30\tofour    &  $<$1\tofour\\
        \hline
6 $b$ tags                    & 2.21\tothree    & 4.97\tothree      &  3.80\tothree   & 8.01\tothree   & 9.57\tofour     & 5.10\tothree    &  1.86\tofour   &  $<$1\tofour \\
$2 \times h\to b\bar b$              & 1.81\tothree    & 5.94\tothree      &  2.01\tothree  & 5.47\tothree   &  6.60\tofour      &  3.28\tothree    & 1.11\tofour    &  $<$1\tofour \\

        \hline
    \end{tabular}
  \end{center}
  \caption{\label{tab:cutflow} Cut flow for the analysis outlined in
    Sec.~\ref{sec:rec}. Boson decays in the background
    samples are generated fully inclusive.}
\end{table*}

Given the small production cross sections, we focus in the following
on semi-leptonic decays of the final state
top pair, with both Higgs bosons decaying $h\to b\bar b$. We use {\sc
  Sherpa} v2.1.1 with the {\sc{Comix}} matrix element
generator~\cite{sherparefs} to generate signal and background events
for modified trilinear Higgs couplings with SM-like top Yukawa
interactions and normalise to the signal events to the NLO cross
section of Ref.~\cite{Frederix:2014hta}. The signal and background
samples have been generated at purely leading order matched to the
parton shower, with modelling of hadronisation effects and underlying
event. Unstable particles are treated in the narrow width
approximation with the subsequent decays conserving any spin
correlations.

The parton distribution functions used are from
CT10~\cite{Guzzi:2011sv} and the scales are set according to
Ref.~\cite{Hoeche:2009rj}.  The masses and widths of the SM particles
used in the event generation are:
\begin{equation}
\begin{split}
M_Z=91.188~\hbox{GeV}\,,  &\quad\Gamma_Z=2.51~\hbox{GeV}\,, \\
M_W=80.419~\hbox{GeV}\,, &\quad\Gamma_W=2.11~\hbox{GeV}\,, \\ 
M_h=126~\hbox{GeV}\, ,    &\quad\Gamma_h=5.36~\hbox{MeV}\,,  \\ 
M_t=173 ~\hbox{GeV}\,,    &\quad\Gamma_t=1.53~\hbox{GeV} \,.
\end{split}
\end{equation}

\section{$t\bar t hh$ at LHC run 2}
\label{sec:analysis}

\begin{figure}[!b]
  \includegraphics[width=0.45\textwidth]{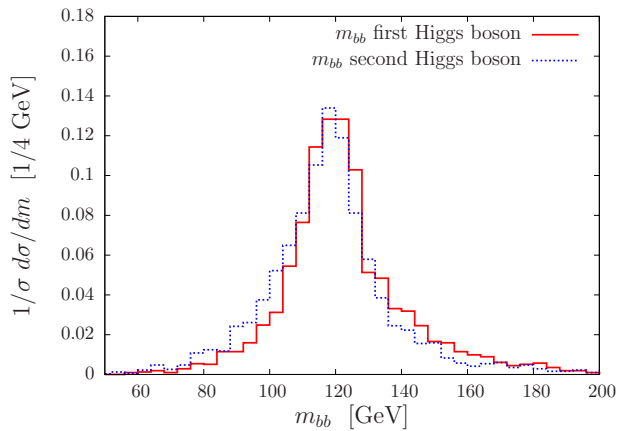}
  \caption{\label{fig:rechiggs} Reconstructed invariant mass of
    bottom-quark pairs based on Eq.~\eqref{eq:mbb1mbb2} for $\lambda =
    \lambda_{\mathrm{SM}}$.}
\end{figure}

\subsection{Final State Reconstruction}
\label{sec:rec}
While this high-multiplicity final state might allow to trigger in
multiple ways, due to the low-$p_T$ thresholds for the jets we rely
for this purpose on an isolated lepton (muon or electron) with
$p_{T,l} > 10$ GeV. We define a lepton to be isolated if the hadronic
energy deposit within a cone of size $R =0.3$ is smaller than $10\%$
of the lepton candidate's transverse momentum and $|y_l |< 2.5$.

After removing the isolated leptons from the list of input particles
($|y|<4.5$) of the jet finder we reconstruct jets with $R=0.4$ and
$p_{T,j}>30$ using the anti-$k_T$ algorithm~\cite{Cacciari:2008gp} of
{\sc{FastJet}}~\cite{fastjet}. We veto events with less than 6
reconstructed jets.

Out of the 6 jets we require at least 5 to be $b$-tagged by matching
the $b$-meson before the decay to the jet. We assume a $b$-tagging
efficiency of $70\%$ and a fake rate of~$1\%$~\cite{atltag}.

As the signal rate after these inclusive cuts is already fairly small,
$\mathcal{O}(10^{-2})$ fb for $\lambda = \lambda_{\text{SM}}$, we
select the Higgs-decay jets by minimising
\begin{equation}
\label{eq:mbb1mbb2}
\chi^2_{HH} = \frac{(m_{b_i,b_j} - m_h)^2}{\Delta_h^2} +
\frac{(m_{b_k,b_l} - m_h)^2}{\Delta_h^2}\,,
\end{equation}
where $k \neq l \neq i \neq j $ run over all $b$-tagged jets and $m_h
= 120$ GeV (we comment on this choice further below) and $\Delta_h =
20$ GeV. For the combination which minimises $\chi^2$ we require
$|m_{b_i,b_j} - m_h| \leq \Delta_h$ and $|m_{b_k,b_l} - m_h| \leq
\Delta_h$.  We then remove these 4 $b$-tagged jets from the event.

To confidently reduce the large gauge boson induced backgrounds,
e.g. $W$+jets, we further require at least one top quark to be
reconstructed. We provide cross sections after the reconstruction of
the leptonic top only, after reconstructing the
hadronic top quark only or after reconstructing either the leptonic or
the hadronic top quark.

To avoid biasing the vector boson backgrounds towards the top quark
signal, for the leptonic top quark reconstruction we require that the
invariant mass of the sum of the lepton, a $b$-jet and the missing
transverse energy vector, built from all visible objects within
$|y|<4.5$, fulfil
\begin{equation}
  |m_{l,b,\slashed{E}} - m_t| \leq \Delta_t \,.
\end{equation}
with $m_t = 170$ GeV and $\Delta_t=40$~GeV.  To identify the $b$-jet
for $m_{l,b,\slashed{E}}$ we consider all remaining $b$-jets in the
event and minimise
\begin{equation}
  \chi^2_{t_l} = \frac{(m_{l,b_i,\slashed{E}} - m_t)^2}{\Delta_t^2}\,.
\end{equation}
Similarly, for the hadronic top quark reconstruction we loop over all
remaining jets and minimise
\begin{equation}
  \chi^2_{t_h} = \frac{(m_{j_i,j_k,j_l} - m_t)^2}{\Delta_t^2}\,.
\end{equation}
We then request
\begin{equation}
|m_{j_i,j_k,j_l} - m_t| \leq \Delta_t \,.
\end{equation}

The cut flow for the described analysis steps is shown in
Tab.~\ref{tab:cutflow}.

\subsection{Discussion}

\begin{figure}[!t]
  \includegraphics[width=0.45\textwidth]{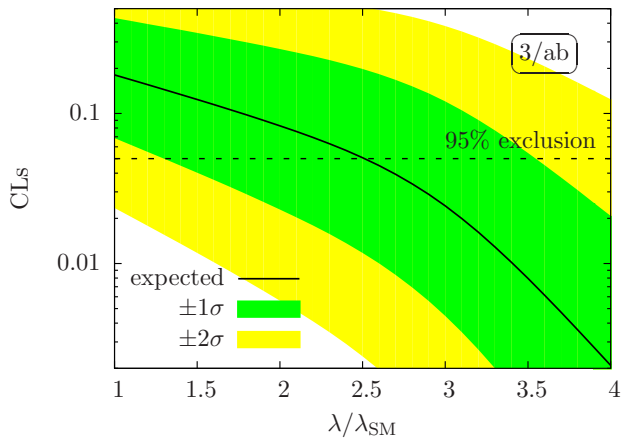}
  \caption{\label{fig:cls} Expected confidence levels for the analysis
    of Sec.~\ref{sec:rec} as a function of the trilinear Higgs
    coupling $\lambda$.}
\end{figure}

At a center-of-mass energy of $14$ TeV, the signal cross section for
$t\bar{t}hh$ is in the sub-femtobarn range before decays are included. 
Therefore, the reconstruction requires an approach that on the one hand 
retains an as large as possible signal yield and on the other hand 
triggers in the high-luminosity regime. We therefore focus on the Higgs 
decays to bottom quarks and semi-leptonic $\bar{t}t$ decays. Other channels can
be combined with the one we focus on to improve the sensitivity on
measuring the self-coupling.

Already after fulfilling the trigger requirement, minimal jet cuts and
5 $b$ tags we find $S/B \simeq 1/15 $ for the backgrounds we
consider. To confirm the measurement of a di-Higgs event both Higgs
bosons have to be fully reconstructed. At this stage we find $S/B
\simeq 1/9$ with 5 $b$ tags and $S/B \simeq 1/6$ with 6 $b$ tags
respectively. We show the reconstructed masses of the hardest and
second hardest Higgs boson in Fig.~\ref{fig:rechiggs}. Due to the
partly invisible decay of B-mesons, $m_H$ is systematically shifted to
slightly lower values. This is why we choose $m_H=120$ GeV for the
minimisation procedure. In measurements, the experiments can
compensate for this systematic shift in the invariant Higgs mass using
$b$-jet calibrations.  Further, at this point with the chosen
b-tagging-efficiency working point $W$+jets backgrounds are already
subleading. Thus, choosing a higher $b$-tagging efficiency working
point at cost of a larger fake rate could be beneficial in this
analysis to retain a larger signal yield and improve the statistical
significance expressed in $S/\sqrt{B}$.

In a further step we then perform a leptonic or hadronic top quark
reconstruction using the remaining measured final state objects. This
can help to further suppress potentially large reducible QCD-induced
backgrounds, e.g. $W$+jets. However, for the top-rich irreducible
backgrounds we focus here mostly on, an improvement in $S/B$ cannot be
achieved using the signal-sparing $\chi^2$ minimisation we apply.

From Tab.~\ref{tab:cutflow} it becomes obvious that the signal
vs. background ratio is expected to be in the $10\%$ range for
$\lambda = \lambda_{\mathrm{SM}}$. After 3/ab we expect 13 signal
events including the reconstruction of a top quark and 22 signal
events reconstructing only the two Higgs bosons. While the signal
yield is too small to claim a discovery at this stage the number of
observed events is high enough to formulate an expected 95\%
confidence level limit on $\lambda$ assuming $y_t=y_t^{\text{SM}}$. In
order to do this, we employ the CLs method~\cite{Read:2002hq,junk}
inputting the expected number of signal and background events for a
luminosity of 3/ab including the reconstruction of at least one top
quark. The result is shown in Fig.~\ref{fig:cls}; and we obtain
\begin{equation}
  \lambda\lesssim 2.51~\lambda_\text{SM}~\hbox{at 95\% CLs.}
\end{equation}

Together with analyses of the $b\bar b \gamma\gamma$ and $b\bar b
\tau\tau$ channels that yield a confidence interval $\lambda \gtrsim
1.3~\lambda_{\text{SM}}$ \cite{uli,us2}, depending on systematic
uncertainties, $t\bar t hh$ will allow us to extend the sensitivity
range and in fact to almost entirely cover the parameter $\lambda$ at
the end of LHC run 2.

\section{Summary and Conclusions}
\label{sec:conc}
With current Higgs property measurements strongly indicating a SM-like
character of the discovered Higgs boson, analysis strategies for
parameters relevant for electroweak symmetry breaking that remain
unconstrained in standard Higgs searches will play a central role in
the search for new physics beyond the SM during run 2. Constraining
the Higgs self-interaction as one of the most interesting couplings in
this regard is a experimentally challenging task and will require a
large accumulated data set.

As we have discussed in this letter, the role of $pp\to t\bar t hh$
production in this regard is twofold: Firstly, it provides an
additional channel that can be added to a global Higgs self-coupling
analysis across the phenomenologically viable channels. Signal
vs. background ratios indicate that top-pair associated Higgs pair
production can provide significant statistical power to increase the
sensitivity to this crucial coupling at a targeted 3/ab and extend the
sensitivity coverage to the Higgs trilinear coupling. Secondly, if we
face a situation with $\lambda\gtrsim \lambda_{\text{SM}}$, $pp\to
t\bar t hh$ provides the {\emph{leading}} channel, where we can expect
to observe an excess over the SM expectation. A negative search
outcome in GF and WBF dominated search strategies in addition to an
excess in $t\bar t hh$ final states would therefore be a strong
indication of $\lambda>\lambda_{\text{SM}}$, eventually allowing us to
put strong constraints on BSM scenarios such as composite Higgs
models.


{\emph{Acknowledgements.}} 
CE is supported by the Institute for Particle
Physics Phenomenology Associateship programme.


\end{document}